\documentclass[%
 reprint,
superscriptaddress,
 amsmath,amssymb,
 aps,
]{revtex4-2}

\usepackage{physics}
\usepackage{graphicx}
\usepackage{dcolumn}
\usepackage{bm}
\usepackage[hidelinks]{hyperref}

\usepackage[danish,spanish,english]{babel}
\usepackage[dvipsnames]{xcolor}

\begin{document}

\preprint{APS/123-QED}

\title{Entanglement properties of a quantum-dot biexciton cascade in a chiral nanophotonic waveguide}

\author{Eva M. González-Ruiz}
\email{eva.ruiz@nbi.ku.dk}
\affiliation{%
 Center for Hybrid Quantum Networks (Hy-Q), Niels Bohr Institute\\
 University of Copenhagen, Blegdamsvej 17, DK-2100 Copenhagen, Denmark
}%

\author{Freja T. {\O}stfeldt}
\affiliation{%
 Center for Hybrid Quantum Networks (Hy-Q), Niels Bohr Institute\\
 University of Copenhagen, Blegdamsvej 17, DK-2100 Copenhagen, Denmark
}%
\author{Ravitej Uppu}
\affiliation{%
Department of Physics \& Astronomy, University of Iowa,
Iowa City, IA 52242
United States
}%
\affiliation{%
 Center for Hybrid Quantum Networks (Hy-Q), Niels Bohr Institute\\
 University of Copenhagen, Blegdamsvej 17, DK-2100 Copenhagen, Denmark
}%
\author{Peter Lodahl} 
\affiliation{%
 Center for Hybrid Quantum Networks (Hy-Q), Niels Bohr Institute\\
 University of Copenhagen, Blegdamsvej 17, DK-2100 Copenhagen, Denmark
}%
\author{Anders S. Sørensen} 
\affiliation{%
 Center for Hybrid Quantum Networks (Hy-Q), Niels Bohr Institute\\
 University of Copenhagen, Blegdamsvej 17, DK-2100 Copenhagen, Denmark
}%

\date{\today}

\begin{abstract}
We analyse the entanglement properties of deterministic path-entangled photonic states generated by coupling the emission of a quantum-dot biexciton cascade to a chiral nanophotonic waveguide, as implemented by \O{}stfeldt et al. [PRX Quantum \textbf{3}, 020363 (2022)]. 
We model the degree of entanglement through the concurrence of the two-photon entangled state in the presence of realistic experimental imperfections. The model accounts for imperfect chiral emitter-photon interactions in the waveguide and the asymmetric coupling of the exciton levels introduced by fine-structure splitting along with time-jitter in the detection of photons. The analysis shows that the approach offers a promising platform for deterministically generating entanglement in integrated nanophotonic systems in the presence of realistic experimental imperfections. 
\end{abstract}

\maketitle




\section{Introduction}
\begin{figure*}
\centering
\includegraphics[width=14cm]{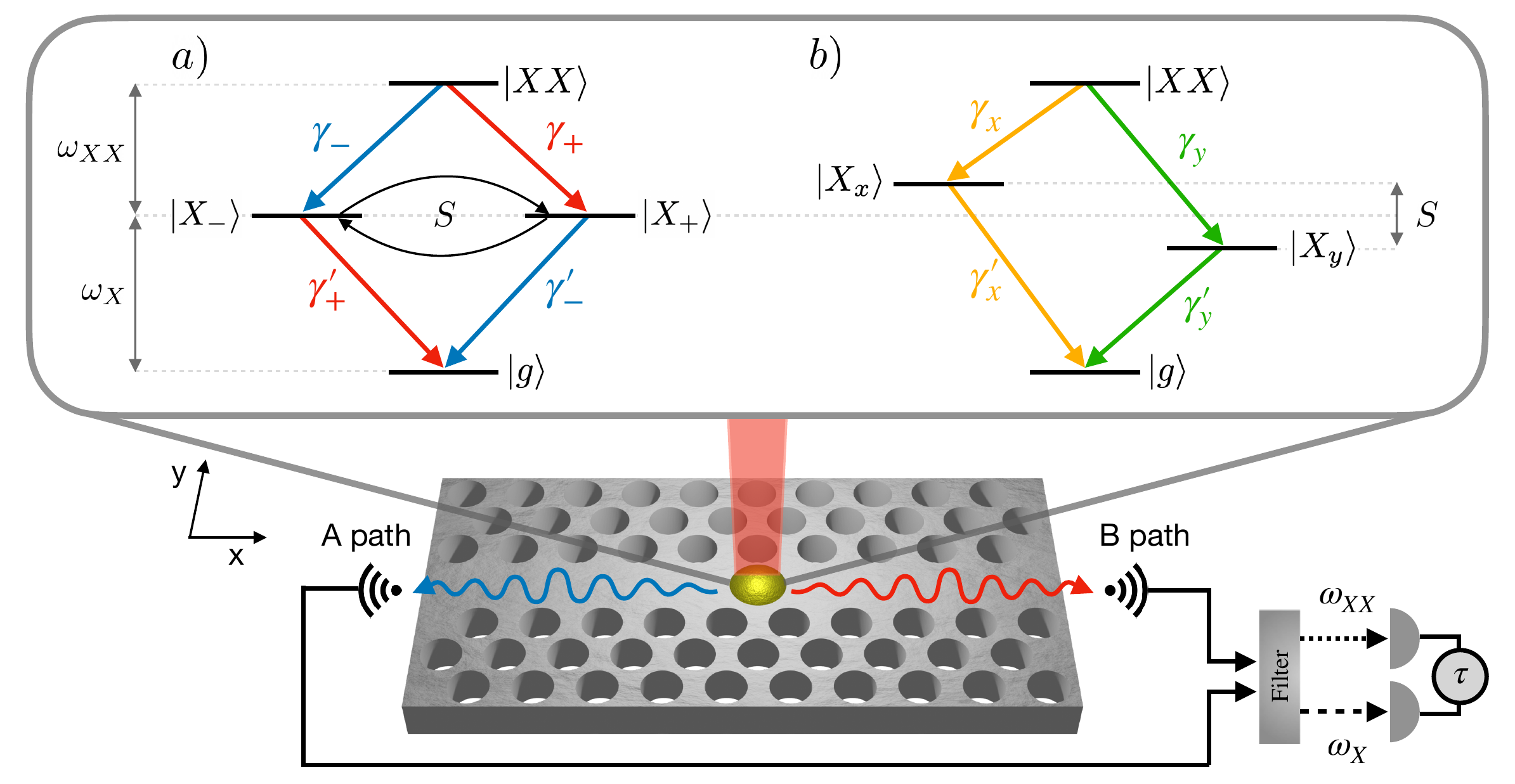}
\caption{\label{fig:scheme} Entanglement generation scheme and level structure of the quantum dot (QD). The QD (semisphere) is placed in a chiral nanowaveguide, and is excited from above, perpendicularly to the nanostructure plane. The photons emitted by the QD can couple to the left (A path) or to the right (B path). The light is collected and frequency-filtered to separate between the biexciton ($\omega_{XX}$) and exciton ($\omega_{X}$) photons in order to measure the desired temporal correlations as a function of $\tau$, the difference between the two emission times. The level structure of the QD can be expressed in two different bases: a) \textit{Circular basis}. The biexciton level $\ket{XX}$, with energy $\hbar\omega_{XX}$, emits two opposite circularly-polarised photons (circular right and left polarised with $\gamma_\pm$ decay rates, respectively). In this picture, the two exciton levels $\ket{X_{+}}$ and $\ket{X_{-}}$ have the same energy $\hbar\omega_{X}$, but are coupled 
at a frequency equal to the fine-structure splitting $S$, which makes the state time-dependent. The exciton $\ket{X_{+}}$ decays at rate $\gamma'_-$ to the ground state $\ket{g}$, and so does $\ket{X_{-}}$ at a rate $\gamma'_+$. b) \textit{Linear basis}. The biexciton level $\ket{XX}$ has the same energy as in the circular basis, but the two emitted photons have opposite linear polarisation (horizontal and vertical with $\gamma_x$ and $\gamma_y$ decay rates, respectively). The two exciton levels are no longer degenerate, but split into the exciton levels $\ket{X_x}$ and $\ket{X_y}$, that are stationary in time. The exciton level $\ket{X_x}$ ($\ket{X_y}$) couples to the $x$ ($y$) in-plane dipole component and has energy $\omega_x+S/2$ ($\omega_x-S/2$). It decays to the ground state $\ket{g}$ at a rate $\gamma'_x$ ($\gamma'_y$). }
\end{figure*}

The generation of high-fidelity entanglement is key for the development of modern quantum technologies \cite{horodecki2009,jozsa2003}. Entangled states of photons have been widely generated probabilistically by employing spontaneous parametric down-conversion (SPDC) \cite{zhong2018}, but the probabilistic nature of this process is a major obstacle for scaling up to high photon numbers. The possibility of entanglement generation on demand is of utmost importance for a wide range of quantum information applications, such as measurement-based quantum computing \cite{bartolucci2021,briegel2009}. The biexciton cascade from quantum-dot (QD) photon sources has been investigated as an on-demand entanglement generator \cite{benson2000,akopian2006,liu2019,huber2018}. The emitted states are, however, entangled in the polarisation degree of freedom, which is incompatible with implementations in integrated photonic circuits \cite{politi2009} that typically support only a single polarisation mode. This poses a challenge for future integration and scalability of quantum technologies \cite{wang2020a} relying on biexciton-cascade entanglement sources.

A solution to the integration of the biexciton source into nanophotonic devices was presented in Ref.~\cite{freja}. Here the photon emission from a cascaded-biexciton decay from InGaAs quantum dots was coupled to a chiral nanophotonic waveguide \cite{sollner2015} (see Fig.~\ref{fig:scheme}). The polarisation-dependent directional emission enabled by chiral coupling of dipoles in these waveguides enable a promising route for on-chip, path-entangled photon generation. Two-photon excitation of the quantum dot  prepares the system in the biexciton state $\ket{XX}$ with energy $\omega_{XX}+\omega_{X}$, which decays through two possible channels to the exciton levels $\ket{X_{\pm}}$ (see Fig.~\ref{fig:scheme}(a)). In a homogenous medium, the biexciton decays radiatively to one of the exciton levels, emitting a photon with either right ($\sigma_+$) or left ($\sigma_-$) circular polarisation. The two exciton levels, $|X_+\rangle$ and $|X_-\rangle$, are degenerate with energy $\omega_{X}$ and decay to the ground state $\ket{g}$ emitting photons with opposite circular polarisation to that emitted during the biexciton decay due to angular momentum conservation. The two emitted photons are thus entangled in polarisation as there is no information regarding which decay path the system followed.
To turn this into a chip-compatible, path-entangled photon source, the QD is placed in a single-mode chiral photonic crystal waveguide which allows converting the polarisation of the transition dipole moment to the emission direction of the photon, i.e. $\sigma_-$ dipoles emit to the left (path A) and $\sigma_+$ dipoles emit to the right (path B). The polarisation entangled state created by the biexciton cascade is thus translated into path encoding that can be used in integrated photonic circuits. Ref.~\cite{freja} 
reported on experimental measurements of the dynamics by out-coupling the photons from the waveguide and frequency-filtering them in order to separate photons emitted on the biexciton and exciton transitions. The desired correlations were then measured through a Hanbury-Brown-Twiss (HBT) experiment \cite{HBT}, as shown in Fig.~\ref{fig:scheme}.
\enlargethispage{-7cm}

While an ideal QD that is precisely positioned at a chiral point could generate maximally entangled, path-encoded photon pairs, imperfections in the QD as well as in the chiral coupling could impact the degree of entanglement. In particular, intrinsic asymmetry of the QD could lead to coupling between the exciton states $\ket{X_{\pm}}$ through a spin-flip oscillation with a frequency $S$ that is known as the fine-structure splitting (FSS) of the QD. In this work we provide a full theoretical analysis of the entanglement properties of the path-entangled state accounting for all these imperfections \footnote{The codes used in this study are available at the Electronic Research Data Archive (ERDA) of the University of Copenhagen. DOI: \url{https://doi.org/10.17894/ucph.a68d50d5-9f9b-4c3b-befd-ef99fcfe2959}}. This analysis already successfully described the experimental findings in Ref.~\cite{freja}, but here we provide the full details of the theory and apply it to systematically analyse the impact of various errors on the degree of entanglement. In particular, the aforementioned FSS induces a frequency splitting of the exciton levels (see Fig.~\ref{fig:scheme}), which effectively creates a time dependence of the entangled polarisation states. 
This can reduce the quality of entanglement when imperfect time detection of photons is taken into account. Moreover, since the photons emitted in the two different decay paths in Fig.~\ref{fig:scheme}(b) have different polarisations, the two paths may occur with different probabilities in photonic nanostructures, given by the polarisation dependent local density of states. 
 These effects, together with imperfect chiral coupling to the waveguide, can reduce the amount of entanglement. The analysis and understanding of these effects will be important for further explorations of the biexciton cascade as an on-demand source of path-entangled photons in integrated quantum information platforms.
\enlargethispage{-3cm}
\section{Analysis}
We start our analysis by introducing the Hamiltonian of the system and a wavefunction ansatz for the state generated by means of the light-mater interaction with the QD.
The state is then fully characterised through studying its evolution by solving Schr\"odinger's equation. 

\subsection{The Hamiltonian and wavefunction ansatz}
The biexciton level structure can be expressed in two different bases. In the linear polarisation basis (Fig.~\ref{fig:scheme}(b)), the emitted photons are linearly polarised (either horizontally or vertically, with $\gamma_x$ and $\gamma_y$ decay rates, respectively), while in the circular basis (Fig.~\ref{fig:scheme}(a)) the photons are circularly polarised (with right- and left-circularly polarised photons, and $\gamma_+$ and $\gamma_-$ decay rates, respectively). In the linear basis the two exciton levels have different energies, split by the FSS $S$, while in the circular basis the levels are degenerate. In the latter basis, there is a time-dependent oscillation between the two exciton levels at a frequency $S$.

The full system is described by the total Hamiltonian $\hat{H} = \hat{H}_0 + \hat{H}_\textrm{f} + \hat{H}_\textrm{int}$, which can be decomposed into the free energy of the emitter- $\hat{H}_0$, the free field- $\hat{H}_\textrm{f}$ and the interaction $\hat{H}_\textrm{int}$ Hamiltonians. These are given by
\begin{align}
  \begin{split}
    \hat{H}_{0} &= \hbar\left( \omega_{XX} + \omega_{X} \right) \ket{XX}\bra{XX} \\
    & + \hbar\left( \omega_{X} + \frac{S}{2} \right) \ket{X_x}\bra{X_x} + \hbar\left( \omega_{X} - \frac{S}{2} \right) \ket{X_y}\bra{X_y} \\
    \hat{H}_\textrm{f} &= \hbar\int\bigg(\omega_{\mathbf{k}}\hat{a}^{\dagger}_{\mathbf{k}}\hat{a}_{\mathbf{k}} + \omega'_{\mathbf{k}}\hat{a}'^{\dagger}_{\mathbf{k}}\hat{a}'_{\mathbf{k}}\bigg)d\mathbf{k} \\
    \hat{H}_\textrm{int} &= -\frac{q}{m_0} \hat{\boldsymbol{A}}\cdot\hat{\boldsymbol{p}}\,,
    \label{eq:Hamiltonian}
  \end{split}
\end{align}
where we have chosen 
the Coulomb gauge with vector potential $\mathbf{A}$. The QD is described by the coordinate $\mathbf{r}$ with the conjugate variable or generalised momentum $\mathbf{p}$, charge $q$ and mass $m_0$ \cite{lodahl_review}. 
Ideally, the energy of the biexciton ($\ket{XX}$) and exciton ($\ket{X_{\alpha}}$ with $\alpha=x,y$) levels is given by $\hbar\omega_{XX}$ and $\hbar\omega_{X}$, respectively. The FSS $S$, however, splits the exciton levels into $\hbar(\omega_{X}\pm S/2)$ in the linear polarisation basis. Note that we express the total Hamiltonian in a linear polarisation basis as it simplifies the temporal dynamics of the system. The field annihilation operators $\hat{a}_{\mathbf{k}}$ are momentum dependent, where $\mathbf{k}$ expresses the corresponding wavevector,
and the prime indicates whether it annihilates a biexciton ($\hat{a}_{\mathbf{k}}$) or an exciton ($\hat{a}'_{\mathbf{k}}$) photon with frequency $\omega^{(\prime)}_{\mathbf{k}}$, correspondingly. The biexciton and exciton binding energies are assumed to be sufficiently different to treat them as two independent reservoirs. This assumption is motivated by the 2 -- 3 meV energy splitting between the exciton and biexciton binding energies observed in QDs, which is over three orders of magnitude larger than the natural linewidths of these transitions \cite{pedersen2020}.

To put the interaction Hamiltonian into a simpler form, the conjugate variable $\mathbf{p}$ (proportional to the dipole operator) can be expressed in terms of the transition matrix elements $\mathbf{\hat{p}} = \sum_{l,m} \bra{l}\mathbf{\hat{p}}\ket{m}\ket{l}\bra{m}$, where the indexes $l$ and $m$ represent the excited and ground states of the transition, respectively. This allows us to express the interaction Hamiltonian as 
\begin{equation}
  \hat{H}_\textrm{int} = \sum_{l,m,\mathbf{k}} \bra{l}\mathbf{\hat{p}}\ket{m} \cdot \mathbf{U}_{\mathbf{k}(\mathbf{r})}\hat{a}_k\ket{l}\bra{m}+{\rm H.c.}\,,
\end{equation}
where $\mathbf{U}_{\mathbf{k}(\mathbf{r})}$ is the mode-function of the field. We consider that the field propagates in the waveguide along the $x$ direction. Following Bloch's theorem we thus have $ \mathbf{U}_{k}(\mathbf{r}) = \mathbf{e}_{k}(\mathbf{r}) e^{ikx}$, where $\mathbf{e}_{k}(\mathbf{r})$
is the Bloch function describing the electric field with wavenumber $k$ at the QD position $\mathbf{r}$, and the field only propagates in the $x$ direction. Moreover we assume that the QD only interacts within a narrow frequency range around the resonance frequency with wavenumbers $\pm k_0$ yielding
\begin{equation}
  \hat{H}_\textrm{int} = \sum_{\substack{l,m\\k\approx \pm k_0}} \bra{l}\mathbf{\hat{p}}\ket{m} \cdot\mathbf{e}_{k}(\mathbf{r})e^{ikx}\hat{a}_{k}\ket{l}\bra{m}+{\rm H.c.}\,,
  \label{eq:int_ham}
\end{equation}
where for brevity we have taken only the non primed annihilation operators, with the sign of $k$ indicating whether the field propagates to the right ($+k_0$) or to the left ($-k_0$). By assuming the same wavenumber in both directions, we implicitly assume time-reversal symmetry for the propagation of the field in the waveguide (i.e. without the QDs). This is valid as long as we can e.g. neglect the intrinsic Faraday effect of the waveguide. Since waveguides are very broad band this is typically an excellent approximation 
and does not exclude any possible violation of time-reversal symmetry of the QD if an external magnetic field was applied.


The polarisation of the emitted light is determined by the symmetry of the states, which results in the following matrix elements for the dipole forbidden transitions in the linear polarisation basis
\begin{equation}
\begin{split}
   \bra{XX}\hat{p}_x\ket{X_y} &= \bra{XX}\hat{p}_y\ket{X_x} \\
   &=\bra{X_x}\hat{p}_y\ket{g} = \bra{X_y}\hat{p}_x\ket{g} = 0\,, 
\end{split}
\end{equation}
as the $x$ ($y$) component of the dipole only couples to the horizontally (vertically) polarised light. Moreover, the allowed transitions from the exciton levels have a dipole moment defined as $P$,
\begin{equation}
  \bra{X_x}\hat{p}_x\ket{g} = \bra{X_y}\hat{p}_y\ket{g} = P \,,
\end{equation}
whereas the two possible biexciton decay transitions are given by \cite{lodahl_review}
\begin{equation}
\bra{XX}\hat{p}_x\ket{X_x} = \bra{XX}\hat{p}_y\ket{X_y} = \sqrt{2}P\,.
\label{eq:biexciton_decay}
\end{equation}
We now insert these dipole transitions in the interaction Hamiltonian from Eq.~\eqref{eq:int_ham} and calculate its Fourier transform. 
For now we only consider the modes propagating to the right (path B), yielding
\begin{widetext}
\begin{equation}
\begin{aligned}
  \hat{H}_\textrm{int} 
  = -P\cdot
  \bigg[\sqrt{2} &\bigg( \epsilon_{k_0,x}(\mathbf{r})\ket{XX}\bra{X_x}+\epsilon_{k_0,y}(\mathbf{r})\ket{XX}\bra{X_y}\bigg)e^{ik_0x_0}\hat{a}_B(x_0) \\
  +&\bigg( \epsilon_{k_0',x}(\mathbf{r})\ket{X_x}\bra{g}+ \epsilon_{k_0',y}(\mathbf{r})\ket{X_y}\bra{g}\bigg)e^{ik_0x_0}\hat{a}_B'(x_0) \bigg] + \text{H.c.},
  \end{aligned}
  \label{eq:ham_int_fourier}
\end{equation}
\end{widetext}
where the position-dependent annihilation operator $\hat{a}_n(x)$ is defined as
\begin{equation}
  \hat{a}_n(x) = \frac{1}{\sqrt{2\pi}}\int_0^\infty  \hat{a}_{n,\pm k}e^{i(k-k_0)x}dk\,,
\label{eq:a_pos}
\end{equation}
with $n=B(A)$ denoting fields propagating to the right (left) and the sign being positive (negative) for path $B$ ($A$) and $x_0$ is the position of the emitter. We note that since we separate the annihilation operator into left and right propagating modes ($A$ and $B$) the limit of the integration is $k=0$. In practice, however, we only expect the annihilation operator to give a contribution for $k\approx \pm k_0 $. We can therefore extend the limit of integration to $-\infty$ yielding the commutator 
\begin{equation}
 [ \hat{a}_n(x) , \hat{a}^{\dagger}_{n'}(x') ]=\delta_{n,n'}\delta(x-x')\,.
\end{equation}
We further note that with the definition in Eq. \eqref{eq:a_pos} we make the convention that both left and right propagating fields are traveling towards positive $x$, i.e. the direction of the $x$-axis is reversed for the left propagating modes. 

To relate the coupling of the right propagating modes with the left propagating modes we again invoke time-reversal symmetry of the waveguide modes. If the local electric field $\mathbf{\epsilon}_{k_0,x}(\mathbf{r})$ is a solution for the waveguide, then by time-reversal symmetry 
the solution for a wave propagating in the opposite direction is given by
$\mathbf{\epsilon}_{-k_0}(\mathbf{r}) =\mathbf{\epsilon}^{*}_{k_0}(\mathbf{r})$. 
This allows us to obtain the full interaction Hamiltonian by combining Eq.~\eqref{eq:ham_int_fourier} with the corresponding expression for back-propagating waves. This results in
\begin{equation}
\begin{aligned}
  \hat{H}_\textrm{int} &= -\hbar\sum_{\alpha} \bigg[\bigg(g_{ A,\alpha} \hat{a}_{A}(0) +g_{ B,\alpha} \hat{a}_{B}(0)\bigg) \ket{XX}\bra{X_ \alpha} \\
  & \quad + \bigg(g'_{ A,\alpha}\hat{a}'_{A}(0) +g'_{ B,\alpha}\hat{a}'_{B}(0)\bigg) \ket{X_ \alpha}\bra{g} +\text{H.c.} \bigg]\,,
  \label{eq:int_ham_2}
  \end{aligned}
\end{equation}
where we have set $x_0=0$ for simplicity and defined the complex coupling constants $g_{n,\alpha}=|g_{n,\alpha,n}| 
e^{i\phi_{n,\alpha}}$ and their phases in relation to the local electric field components $\epsilon_{\pm k_0,i}$ as
\begin{equation}
  \begin{aligned}[t]
    g_{A,x} &= \sqrt{2}P\epsilon^{*}_{k_0,x}(\mathbf{r}), \\
    g_{B,x} &= \sqrt{2}P\epsilon_{k_0,x}(\mathbf{r}), \\
    g'_{A,x} &= P\epsilon'^{*}_{k_0,x}(\mathbf{r}),\\
    g'_{B,x} &= P\epsilon'_{k_0,x}(\mathbf{r})\,.
  \end{aligned}
  \qquad
  \begin{aligned}[t]
    g_{A,y} &=\sqrt{2}P\epsilon^{*}_{k_0,y}(\mathbf{r}), \\
    g_{B,y} &= \sqrt{2}P\epsilon_{k_0,y}(\mathbf{r}), \\
    g'_{A,y} &= P\epsilon'^{*}_{k_0,y}(\mathbf{r}), \\
    g'_{B,y} &=P\epsilon'_{k_0,y}(\mathbf{r})\,.
  \end{aligned}
  \label{eq:map_g_e}
\end{equation}

The coupling constants in Eq.~\eqref{eq:map_g_e} describe the light-matter interaction between the field and the waveguide including the chirality. In particular, their magnitude describes the coupling of a horizontally or vertically polarised photon (through the $x$ and $y$ components of the dipole, respectively) to the left or to the right paths. From Eq.~\eqref{eq:map_g_e} we note that $|g_{A,\alpha}|=|g_{B,\alpha}|$ for $\alpha=x,y$, so that linearly polarized dipoles always have the same coupling constant and hence the same decay rate in both directions $A$ and $B$. This does not, however, exclude that circular dipoles can have chiral interaction and predominantly decay in one direction. The existence of such chiral interactions is encoded in the relative phase of the coupling constants. 
From Eq.~\eqref{eq:map_g_e} we find that the phase difference $\Phi$ between the phases of the $x$ and $y$ components of the electric field
is
\begin{equation}
  \Phi \equiv \phi_{x} - \phi_{y} = \phi_{A,x} - \phi_{A,y} = - \left( \phi_{B,x} - \phi_{B,y} \right)\,,
  \label{eq:phase_eq}
\end{equation}
and similarly for the exciton phase difference $\Phi'$. Consider now a circularly polarized state $\ket{X_\pm}=(\ket{X_x}\pm i\ket{X_y})/\sqrt{2}$. We can calculate the coupling constants $g'_{n,+}$ for the decay of these states into the $n=A,B$ directions from the interaction Hamiltonian~\eqref{eq:int_ham_2}, yielding
\begin{equation}
    g'_{n,\pm} = \frac{1}{\sqrt{2}}(g'_{n,x}\mp ig'_{n,y})\,.  
\end{equation}
If $|g'_{n,x}|=|g'_{n,y}|=g'$, the decay rate of the circular states into the two directions will thus fulfill
\begin{equation}
\begin{aligned}
    \gamma'_{A,\pm}\propto  |g'_{A,\pm}|^2 = 
    {g'}^2(1\pm \sin\Phi')\\
    \gamma'_{B,\pm}\propto  |g'_{B,\pm}|^2 = 
    {g'}^2(1\mp \sin\Phi')\,.
\end{aligned}    
\end{equation}
For $\Phi' = \pi/2$ the $x$ and $y$ components of the field in the waveguide are phase-shifted corresponding to circular polarisation. Furthermore, whether the waveguide mode is left- or right-hand circularly polarized is linked to the propagation direction of the light. As a consequence, the system exhibits perfect chiral coupling with the circularly polarized states coupling only to a single propagation direction, i.e. $\gamma'_{A,+}\neq 0$ and $\gamma'_{B,+}=0$, with the directions reversed for the opposite circular state. 
Complete absence of chirality occurs when $\Phi' = 0$, where the field in the waveguide is linearly polzarized. 
Thus, the parameters $\Phi$ and $\Phi'$ represent the degree of chirality of the system, which we employ in the subsequent sections of this article. 

To describe the emission into the waveguide, it is convenient to change the Hamiltonian into the position basis. While the Fourier transform of the free energy term in the total Hamiltonian~\eqref{eq:Hamiltonian} is itself, Fourier transforming the free field term 
yields
\begin{align}
  \begin{split}
    &\hat{H}_\textrm{f} = \sum_{n}\bigg[i\hbar\int \bigg(v_{gXX}\frac{\partial \hat{a}^{\dagger}_{n}(x)}{\partial x}\hat{a}_{n}(x) \\
    &\quad \quad \quad \quad \quad \quad \quad + v_{gX} \frac{\partial \hat{a}'^{\dagger}_{n}(x)}{\partial x}\hat{a}'_{n}(x)\bigg)dx\\
    &\quad + \hbar\int\bigg(\omega_{XX}\hat{a}^{\dagger}_{n,k}\hat{a}_{n,k} + \omega_{X}\hat{a}'^{\dagger}_{n,k}\hat{a}'_{n,k}\bigg)dk\bigg]\,,
    \label{eq:ham_field_fourier}
  \end{split}
\end{align}
where the group velocities associated with the biexciton and exciton energy levels are given by $v_{g,XX} = \partial \omega'_{k}/\partial k$ and $v_{g,X} = \partial \omega_{k}/\partial k$ respectively. Note that these two group velocities could be different due to the dispersion of the waveguide and the different emission wavelengths of the exciton and biexciton levels. Here, we approximate them to lowest order around the exciton and biexciton frequencies, that is $\omega_{k} \approx \omega_{X} + v_{g,X}(k-k_0)$ and $\omega'_{k} \approx \omega_{XX} + v_{g,XX}(k-k_0)$.

We can now write a wavefunction ansatz for the total state of the system in the real space domain. The state should describe that up to two photons can be emitted by the biexciton decay and that they couple into the left- or right-propagating waveguide modes. 
Based on the methods from Ref. \cite{sumanta} (with similar methods being developed in Refs.~\cite{fischer2018,trivedi2018,heuck2020}) we use the following ansatz:
\begin{widetext}
\begin{align}
  \begin{split}
    \ket{\psi(t)} = e^{-i(\omega_{XX} + \omega_{X})t}&(c_{XX}(t)\ket{XX}\ket{\emptyset}    +\sqrt{v_{gXX}}\sum_{\alpha,n}\int dt_{XX}\psi_{\alpha,n}(t,t_{XX})\hat{a}^{\dagger}_{n}(v_{gXX}(t-t_{XX}))\ket{X_\alpha}\ket{\emptyset} \\
    &+\sqrt{v_{gXX}v_{gX}}\sum_{n,m}\iint dt_{XX}dt_{X}\psi_{n,m}(t,t_{XX},t_{X})\hat{a}^{\dagger}_{n}(v_{gXX}(t-t_{XX}))\hat{a}'^{\dagger}_{m}(v_{gX}(t-t_{X}))\ket{g}\ket{\emptyset})\,,
    \label{eq:ansatz}
  \end{split}
\end{align}
\end{widetext}
where $t_{X}$ and $t_{XX}$ are the two emission times with $t_{XX}<t_{X}$. This state describes that with an amplitude $c_{XX}(t)$ the system is in the biexciton state with the field being in the vacuum state $\ket{\emptyset}$. Since the system is initially excited to this state we have $c_{XX}(t=0)=1$. The amplitude $\psi_{\alpha,n}(t,t_{XX})$ describes the state after the emission of a photon in the direction $n=A,B$ at time $t_{XX}$ by the decay into the exciton state $\ket{X_\alpha}$. Since the photon propagates in the waveguide, this is associated with a photon at position $x=v_{gXX}(t-t_{XX})$. As this state still evolves in time the amplitude has an explicit dependence on time $t$ with the amplitude vanishing before the emission, $\psi_{\alpha,n}(t,t_{XX})=0$ if $t\leq t_{XX}$. Finally, after the emission of both photons, the system is in the ground state $\ket{g}$ and the two photons are emitted in directions $n,m$ with amplitude $\psi_{n,m}(t,t_{XX},t_{X})$. This amplitude vanishes unless $t\geq t_{X}\geq t_{XX}$. 
It should be noted that both for the left and right propagation directions in the waveguide $x\in[0,\infty]$, i.e. the reference frame is placed such that in both directions $x$ is positive after the QD.

\subsection{Solving the Schr\"odinger equation}
The wavefunctions $\ket{\psi(t)} $ from Eq.~\eqref{eq:ansatz} should be calculated to describe the state. We thus apply Schr\"odinger's equation $i\hbar\partial \ket{\psi}/\partial t = \hat{H}\ket{\psi}$ to the wavefunction ansatz using the space-domain Hamiltonian. Following the procedure from Ref.~\cite{sumanta}, we obtain the set of coupled differential equations:
\begin{multline}
    \dot{c}_{XX}(t) = -\frac{i}{\sqrt{v_{gXX}}\hbar}\sum_{\alpha,n}g_{\alpha,n}\psi_{\alpha,n}(t,t), \\
    \dot{\psi}_{x,n}(t,t_{XX}) = \frac{iS}{2\hbar}\psi_{x,n}(t,t_{XX}) - \frac{ig^{*}_{x,n}c_{XX}(t)}{\sqrt{v_{gXX}}\hbar}\delta(t-t_{XX}), \\
    - \frac{i}{\sqrt{v_{gX}}\hbar}\sum_{m}g'_{x,n}\psi_{n,m}(t,t_{XX},t), \\
    \dot{\psi}_{y,n}(t,t_{XX}) = -\frac{iS}{2\hbar}\psi_{y,n}(t,t_{XX}) - \frac{ig^{*}_{y,n}c_{XX}(t)}{\sqrt{v_{gXX}}\hbar}\delta(t-t_{XX}) \\
    - \frac{i}{\sqrt{v_{gX}}\hbar}\sum_{m}g'_{y,n}\psi_{n,m}(t,t_{XX},t), \\
    \dot{\psi}_{n,m}(t,t_{XX},t_{X}) =
    - \frac{i}{\sqrt{v_{gX}}\hbar}\sum_{\alpha}g^{\prime *}_{\alpha,n}\psi_{\alpha,n}(t,t_{XX})\delta(t-t_{X}) \,.
  \label{eq:dif_equations}
\end{multline}
We then apply the Laplace transform to the nine equations in Eq.~\eqref{eq:dif_equations}, with the system initially prepared in the biexciton state ($c_{XX}(t=0)=1$). The Laplace transform simplifies solving the coupled differential equations to solving an algebraic problem, where the initial conditions of the system are already specified in the Laplace space instead of in the solution of the differential equations.
Inverting the Laplace transformation now yields
\begin{align}
  \begin{split}
    &\dot{\psi}_{x,n}(t,t_{XX}) = - \frac{ig^{*}_{x,n}c_{XX}(t)}{\sqrt{v_{gXX}}\hbar}\delta(t-t_{XX}) \\
    &\quad- \left(\frac{-iS + \gamma'_{x}}{2\hbar}\right) \psi_{x,n}(t,t_{XX}) -\frac{\Gamma}{2\hbar}\psi_{y,n}(t,t_{XX}) \\
    &\dot{\psi}_{y,n}(t,t_{XX}) = - \frac{ig^{*}_{y,n}c_{XX}(t)}{\sqrt{v_{gXX}}\hbar}\delta(t-t_{XX}) \\
    &\quad- \left(\frac{iS + \gamma'_{y}}{2\hbar}\right) \psi_{y,n}(t,t_{XX}) -\frac{\Gamma^{*}}{2\hbar}\psi_{x,n}(t,t_{XX}) \,,
  \end{split}
  \label{eq:diff2}
\end{align}
with the spontaneous emission rates given by 
\begin{align}
\begin{split}
  \gamma^{ ( \prime )}_{\alpha} &= \sum_n\gamma^{ ( \prime )}_{\alpha,n} \equiv \sum_n\frac{|g^{ ( \prime )}_{\alpha,n}|^2}{v_{gX}}\,.
\end{split}
\end{align}
A coupling between the $\ket{X_x}$ and $\ket{X_y}$ states 
mediated by the local electric field of the waveguide is captured by the cross terms with coupling coefficient 
\begin{align}
  \Gamma = \frac{g'_{A,x}g'^{*}_{A,y}+g'_{B,x}g'^{*}_{B,y}}{v_{gX}}\,,
\end{align}
which is real due to time-reversal symmetry~\eqref{eq:map_g_e}. This coupling is important if e.g. the local electric field in the waveguide is diagonally polarized, which leads to 
$\Gamma=\gamma'_x=\gamma'_y$. 

When solving the coupled set of differential equations~\eqref{eq:diff2} it is convenient to work in a basis that diagonalizes the dynamics, i.e. where the equations decouple. For a rotationally symmetric system, this is the case for any basis, but it is no longer the case once the symmetry is broken. The FSS is induced by the asymmetry of the QD and is assumed to be in the $x$ and $y$-directions such that Eqs.~\eqref{eq:diff2} decouple in that basis. On the other hand, the local waveguide field may have a different orientation, which also breaks the symmetry and thus leads to a coupling between the equations, i.e. $\Gamma\neq 0$.
In practice, however, we typically have $S\gg\Gamma$, e.g. in the experimental implementation in Ref. \cite{freja} the fine structure splitting $S$ was an order of magnitude larger than the exciton emission rate $(\gamma'_x+\gamma'_y)/2$.
The coupling between the exciton levels ($\ket{X_x}$ and $\ket{X_y}$) can therefore be neglected and we set $\Gamma=0$. We note that this assumption may lead to inconsistencies in the obtained results due to incorrect normalization of the state in QDs with small FSS, i.e., $S$ comparable to $(\gamma'_x+\gamma'_y)/2$. 
In the subsequent sections, we use $S=4(\gamma'_x+\gamma'_y)/2$ for which we find that the magnitude differs from unity by $<$6\%. 

We now solve the two coupled differential equations from Eq.~\eqref{eq:diff2} by taking the aforementioned limit $\Gamma=0$, such that the equations decouple. 
We can then straightforwardly solve them by again applying the Laplace transform, obtaining
\begin{widetext}
\begin{align}
  \begin{split}
    c_{XX}(t) &= e^{-\frac{1}{2\hbar}(\gamma_{x}+\gamma_y)t} \\
    \psi_{x,n}(t,t_{XX}) &= -i\sqrt{\gamma_{x,n}} e^{-\frac{1}{2\hbar}(\gamma_{x}+\gamma_y)t_{XX}-\frac{1}{2\hbar}\left(\gamma'_{x}+iS\right)\left(t-t_{XX}\right)-i\phi_{x,n}}\theta(t-t_{XX}) \\
    \psi_{y,n}(t,t_{XX}) &= -i\sqrt{\gamma_{y,n}} e^{-\frac{1}{2\hbar}(\gamma_{x}+\gamma_y)t_{XX}-\frac{1}{2\hbar}\left(\gamma'_{x}-iS\right)\left(t-t_{XX}\right)-i\phi_{y,n}}\theta(t-t_{XX}) \\
    \psi_{n,m}(t,t_{XX},t_{X}) &=
    - e^{-\frac{1}{2\hbar}(\gamma_{x}+\gamma_y)t_{XX}}\Bigg(\sqrt{\gamma_{x,n}\gamma'_{x,m}}e^{-\frac{1}{2\hbar}\left(\gamma'_{x}+iS\right)\left(t_{X}-t_{XX}\right)-i\left(\phi_{x,n}+\phi'_{x,m}\right)}\\ &\quad \quad \quad \quad \quad \quad \quad \quad+\sqrt{\gamma_{y,n}\gamma'_{y,m}}e^{-\frac{1}{2\hbar}\left(\gamma'_{y}-iS\right)\left(t_{X}-t_{XX}\right)-i\left(\phi_{y,n}+\phi'_{y,m}\right)}\Bigg)\theta(t-t_{X}) \theta(t_{X}-t_{XX})  \,,
  \end{split}
  \label{eq:results1}
\end{align}
\end{widetext}
where $\theta(x)$ is the Heaviside step function, i.e. $\theta(x) = 1$ if $x>0$, and $\theta(x) = 0$ otherwise. 

We now calculate the probability of detecting two photons simultaneously at the output of the waveguide in order to analyse the quality of the entanglement. To do so we correlate the biexciton and exciton photons with a time delay $\tau$ in two different settings: when both are coupled to the forward or back-propagating direction (noted as $A_{X}A_{XX}$ and $B_{X}B_{XX}$ respectively) and when they couple to opposite directions ($A_{X}B_{XX}$ and $B_{X}A_{XX}$):
\begin{align}
\begin{split}
  &P_{n,m}(t,t_{XX},t_{XX}+\tau) = \\ &|v_{gXX}||v_{gX}|\bra{\psi(t)}\hat{a}_{n}^{\dagger}(v_{gXX}t)\hat{a}_{n}(v_{gXX}t) \\
  &\quad \quad \quad \quad \cdot \hat{a}'_{m}{}^{\dagger}\left(v_{gX}(t-\tau)\right)\hat{a}'_{m}\left(v_{gX}(t-\tau)\right)\ket{\psi(t)} \,.
\end{split}
\end{align}
With the wavefunction ansatz Eq.~\eqref{eq:ansatz} and the results from Eq.~\eqref{eq:results1} we obtain
\begin{multline}   
    P_{n,m} = |\psi_{n,m}(t,t_{XX},t_{XX}+\tau)|^2  \\ = e^{-(\gamma_{x}+\gamma_y)t_{XX}}\bigg[\gamma_{x,n}\gamma'_{x,n}e^{-\gamma'_{x}\tau}+\gamma_{y,n}\gamma'_{y,n}e^{-\gamma'_{y}\tau}\\
    +2\sqrt{\gamma_{x,n}\gamma_{y,n}\gamma'_{x,m}\gamma'_{y,m}}e^{-\frac{1}{2}(\gamma'_{x}+\gamma'_{y})\tau} \\
    \cdot\cos\left(S\tau + 
    (\phi_{x,n}-\phi_{y,n})+(\phi'_{x,m}-\phi'_{y,m})\right)\bigg]\\
    \cdot\theta(t-t_{XX}-\tau)\,. 
  \label{eq:probabiliti}
\end{multline}

\subsection{Entanglement generation}

The state produced by the biexciton cascade coupled to the chiral waveguide has two different degrees of freedom: the path followed (to the left, $A$, or to the right, $B$) and the respective times of emission of the biexciton ($t_{XX}$) and exciton ($t_{X}$) photons. We project this state in time space by fixing the two times of detection $t_{X}-t_{XX}\equiv \tau > 0$. 
Note that the 
characteristics of the state produced 
depends only on the time difference $\tau$. 

From our wavefunction ansatz in Eq.~\eqref{eq:ansatz}, we post-select the two-photon emission terms by conditioning on detecting photons at times $t=t_{XX}$ and $t=t_{X}$, thus obtaining the state
\begin{multline} 
  \ket{\psi(\tau)} = \frac{1}{\sqrt{N}}( \psi_{AA}(\tau)\ket{AA} + \psi_{AB}(\tau)\ket{AB} \\
  + \psi_{BA}(\tau)\ket{BA} + \psi_{BB}(\tau)\ket{BB}) \,,
\end{multline}
where, 
\begin{equation}
\begin{split}
    N = &\abs{\psi_{AA}(\tau)}^2 + \abs{\psi_{AB}(\tau)}^2 + \abs{\psi_{BA}(\tau)}^2 + \abs{\psi_{BB}(\tau)}^2\,,
\end{split}
  \label{eq:norm}
\end{equation}
is the normalisation factor. Note that we dropped the explicit subscripts for exciton $X$ and biexciton $XX$ photons on the direction index. Instead, we utilize time-ordered emission in the simplified notation, i.e. the subscript $AB$ should be read as $A_{XX} B_{X}$.

In general the two possible decay channels do not have the same spontaneous emission rates, i.e., 
$\gamma_{x}\neq\gamma_{y}$ due to differences of the local electric field components in the waveguide. However, to achieve a high degree of chirality in the waveguide, the two exciton decay rates have to be similar $\gamma_{x}\approx\gamma_{y}$. This was also the case in the recent experiment in Ref. \cite{freja}. For most of the article we therefore set $\gamma_x=\gamma_y$ and $\gamma'_x=\gamma'_y$, but investigate the influence of differences in the rates in Sec. \ref{sec:asym}. Moreover, the biexciton and exciton spontaneous emission rates are given by
\begin{equation}
  \gamma_{x}+\gamma_{y}\equiv \gamma_{XX}, \quad \gamma'_{x}=\gamma'_{y}\equiv \gamma_X\,.
  \label{eq:rates_relation}
\end{equation}
Since the biexciton decays twice as fast according to Eq.~\eqref{eq:biexciton_decay}, if we assume identical group velocities we have that $\gamma_X=\gamma_{XX}/2$. In the rest of the article, we assume this relation between the spontaneous emission rates.

\begin{figure}
\centering
\includegraphics[width = \linewidth]{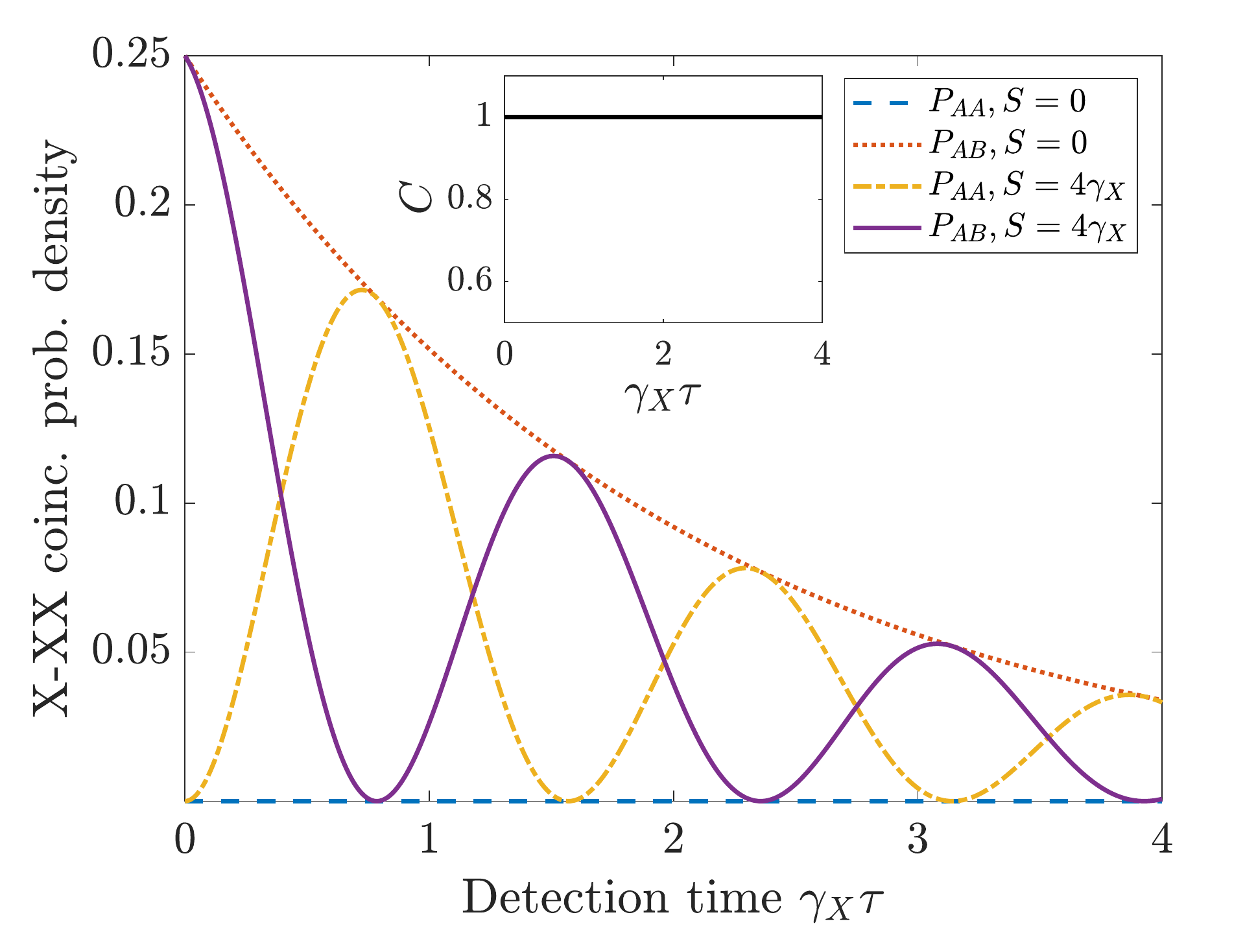}
\caption{\label{fig:S_pi_2} Time correlations $P_{AA}$ and $P_{AB}$ as a function of the difference in the emission time $\tau$. The dashed and dotted lines have been calculated with perfect symmetry between the exciton levels ($S=0$) while the dash-dotted and solid lines have been obtained for $S=4\gamma_X$. The waveguide coupling is perfectly chiral ($\Phi=\pi/2$) in all cases. \textit{Inset}. Concurrence $C$ of the state as a function of the difference in emission times $\tau$. The concurrence remains unity at all times for any value of $S$. This shows that the state is maximally entangled independently of the time $\tau$, as also shown in Eq.~\eqref{eq:pi2_state}.}
\end{figure}

The difference in the phase of the transition dipoles for biexciton and exciton decays, $\Phi$ and $\Phi'$ respectively, satisfies Eq.~\eqref{eq:phase_eq}.
Moreover as the optical wavelengths of the photons emitted from biexciton and the exciton decay channels are comparable, we can approximate the phase differences to be equal, i.e. $\Phi = \Phi'$.
Under these assumptions, the total probability of detecting the first photon at time $t=t_{XX}$ is
\begin{multline}
  P(t=t_{XX}) = (\gamma_x + \gamma_y)e^{-(\gamma_x + \gamma_y)t_{XX}/\hbar} \\
  = 2\gamma_Xe^{-2\gamma_Xt_{XX}/\hbar}\,.
\end{multline}
We can thus calculate the path-dependent, two-photon emission probabilities to be
\begin{align}
  \begin{split}
    P_{AA} &= \frac{\gamma_X}{4}e^{-\gamma_X\tau/\hbar}\left(1 + \cos\left(S\tau+2\Phi\right)\right) \\
    P_{BB} &= \frac{\gamma_X}{4}e^{-\gamma_X\tau/\hbar}\left(1 + \cos\left(S\tau-2\Phi\right)\right)\\ 
    P_{AB} &= P_{BA} = \frac{\gamma_X}{4}e^{-\gamma_X\tau/\hbar}\left(1 + \cos\left(S\tau\right)\right) \,.
  \end{split}
  \label{eq:prob_appr2}
\end{align}
A QD with $S=0$ that is perfectly chiral coupled to the waveguide, i.e., $\Phi = \Phi' =\pi/2$, results in $P_{AA}=P_{BB}=0$.
In this case we thus have the ideal entangled state $(\ket{AB}+\ket{BA})/\sqrt{2}$, where the emission direction of the two photons is perfectly anticorrelated, as shown with the dashed and dotted lines in Fig.~\ref{fig:S_pi_2}. 
Note that, for $S=0$, our model can only accurately represent the perfect chiral coupling case and will lead to erroneous conclusions if $\Phi \neq \pi/2$ since this leads to $\Gamma\neq 0$. 
For the general case of $S > 0$, we can calculate the resulting entangled two-photon state by conditioning the solution in Eq.~\eqref{eq:results1} on the detection of a photon at time $t=t_{XX}$. For perfect chiral coupling the state is
\begin{align}
\begin{split}
  \ket{\psi(\tau)}_{\Phi=\pi/2} = \frac{1}{2}( &\cos\left(\frac{S \tau}{2}\right)\left( \ket{AB} + \ket{BA} \right) \\
  + i &\sin\left(\frac{S \tau}{2}\right)\left( \ket{AA} + \ket{BB} \right) )\,,
  \label{eq:pi2_state}
\end{split}
\end{align}
To understand the entanglement in this state we rewrite it as
\begin{equation}
  \ket{\psi(\tau)}_{\Phi=\pi/2} = \frac{1}{2}(\ket{A}\ket{\xi} + \ket{B}\ket{\xi'})\,,
  \label{eq:entanglement_oscillation}
\end{equation}
which is in fact a maximally entangled state, with $\ket{\xi}=\cos\left(S \tau/2\right)\ket{B}+i\sin\left(S \tau/2\right)\ket{A}$ and $\ket{\xi'}=\cos\left(S \tau/2\right)\ket{A}+i\sin\left(S \tau/2\right)\ket{B}$. For perfect chirality the entanglement is thus maximal regardless of the detection time, although the specific entangled state varies with the emission time, resulting in a time varying detection pattern in Fig.~\ref{fig:S_pi_2}. In practice, this means that the corresponding measurement protocol must compensate for the time dependence. This task may be non-trivial depending on the specific application. Here, for simplicity we chose to characterise the state by its intrinsic entanglement that could be obtained in such an idealized setup. In contrast, if the waveguide interaction is not chiral ($\Phi=0,\pi$) the state is given by 
\begin{align}
\begin{split}
  \ket{\psi(\tau)}_{\Phi=0,\pi} &= \frac{1}{2}\left( \ket{AB} + \ket{BA} + \ket{AA} + \ket{BB} \right) \\
  &= \frac{1}{2}\left( \ket{A} + \ket{B} \right)_X\left( \ket{A} + \ket{B} \right)_{XX}\,,
  \label{eq:phi0_state}
\end{split}
\end{align}
which is a separable state. As a consequence all detection patterns of two photons are equally probable. 
In real experimental settings, the directional (chiral) coupling could lie in between these two extreme cases depending on the local electric field at the location of the QD within the waveguide. This imperfect chirality will lower the entanglement quality of the source, which is quantified in the next section.

\section{Results}

\begin{figure*}
\centering
\includegraphics[width = 16 cm]{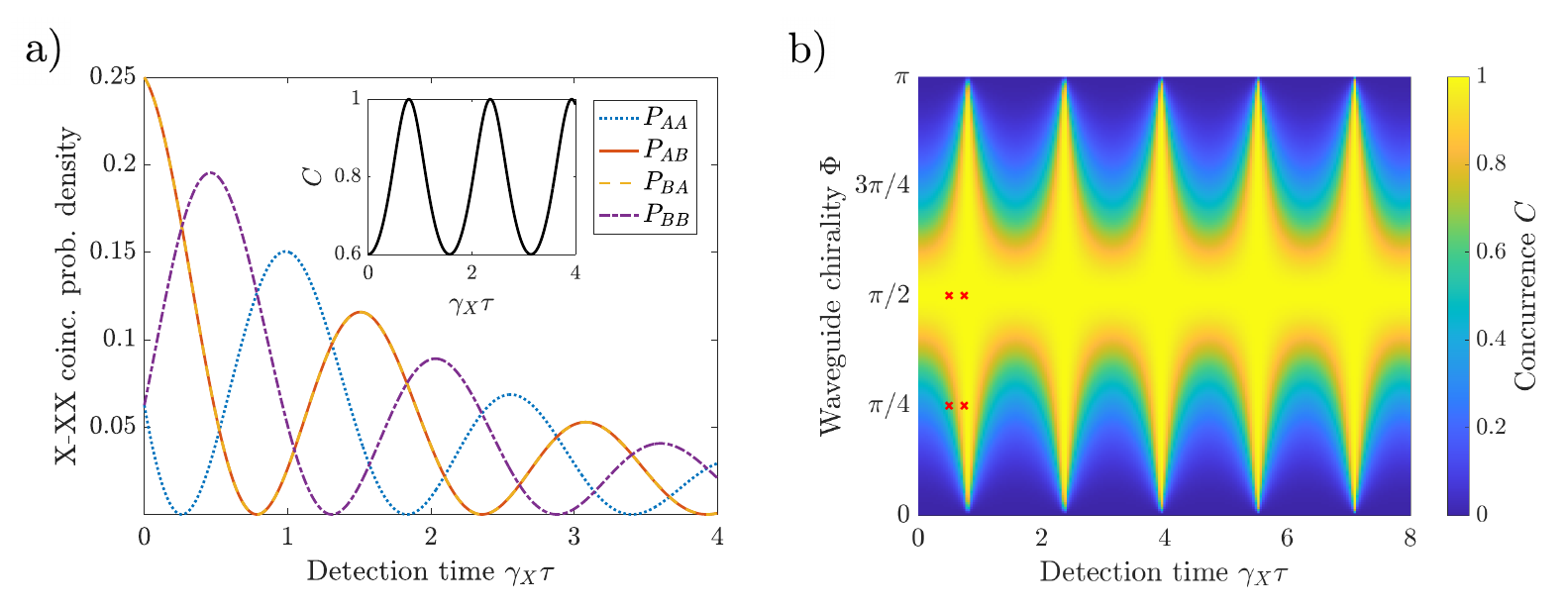}
\caption{\label{fig:3ab} a) Time correlations $P_{AA}$, $P_{AB}$, $P_{BA}$, $P_{BB}$ as a function of the difference in the emission time $\tau$ with a FSS of $S=4\gamma_X$ and waveguide chirality of $\Phi=\pi/3$. We observe that $P_{AA}$ and $P_{BB}$ are out of phase with each other. This may seem surprising, since the two directions are naively the same, but occurs due to an interplay between the imperfect chirality and the sign of the FSS (see text). This effect was experimentally observed in Ref.~\cite{freja}. \textit{Inset}. Concurrence $C$ of the state as a function of the difference in emission times $\tau$. The concurrence oscillates between 0 and 1, as the state evolves. 
b) Colour map of the concurrence $C$ of the state as a function of the difference in phase $\Phi$ and the difference in emission times $\tau$. The concurrence oscillates in time for a given chirality, with the exception of perfect chirality ($\Phi=\pi/2$ which gives $C=1$) and non-chiral waveguide coupling ($\Phi=0,\pi$ which gives $C=0$). The ``$\times$''-markers indicate the points investigated in Figure \ref{fig:time_jitter}(a).}
\end{figure*}

As we have seen in the previous section, the emitted two-photon entangled state depends on the time difference $\tau$ between the biexciton and exciton emission times. We thus expect that any uncertainty in the emission times will affect the entanglement quality of the state. Moreover, imperfect chirality of the waveguide reduces the directionality of emission, thereby leading to non-perfect conversion into path encoding of the entangled state. In this section, we quantify the effect of imperfections on the entanglement quality of the state.

To this end, we employ the concurrence $C$ as the entanglement measure to characterise the quality of the state. The concurrence of any quantum state with a density matrix $\rho$ is given by \cite{concurrence}
\begin{align}
  C(\rho) = \max\{ 0,\lambda_1-\lambda_2-\lambda_3-\lambda_4\}\,,
  \label{eq:concurrence}
\end{align}
where $\{\lambda_i\}$ are the square root of the eigenvalues of $\rho\Tilde{\rho}$ in descending order and $\Tilde{\rho}=(\hat{\sigma}_y\otimes\hat{\sigma}_y)\rho^{*}(\hat{\sigma}_y\otimes\hat{\sigma}_y)$. We calculate the density matrix $\rho$ that represents the path-encoded state obtained from the biexciton cascade to be 
\begin{equation}
  \rho(\tau) = \sum_{\substack{n,n'\\m,m'}}\psi_{n,m}(\tau)\psi^{*}_{n',m'}(\tau) \ket{n,m}\bra{n',m'}\,.
  \label{eq:density_matrix}
\end{equation}
By calculating the resulting eigenvalues $\{\lambda_i\}$, we obtain the concurrence using Eq.~\eqref{eq:concurrence}
\begin{align}
  C(\tau) = \frac{2}{N}\abs{\psi_{AA}(\tau)\psi_{BB}(\tau) - \psi_{AB}(\tau)\psi_{BA}(\tau)}\,.
\end{align}
Inserting the wavefunctions from Eq.~\eqref{eq:results1} and approximating $\gamma_x=\gamma_y$ as discussed earlier (cf. Eq.~\eqref{eq:rates_relation}), the dependence of $C$ on the chiral phase $\Phi$ and the time delay between biexciton and exciton emissions $\tau$ is found to be
\begin{equation}
  C(\Phi,\tau) = \frac{\sin^2\left(\Phi\right)}{1+\cos\left(S\tau\right)\cos^2\left(\Phi\right)}\,.
  \label{eq:concurrence_0}
\end{equation}
We obtain perfect concurrence $C=1$ when the waveguide is perfectly chiral ($\Phi = \pi/2$) as discussed above. 
Furthermore, if the waveguide is completely non-chiral ($\Phi = 0,\pi$) the concurrence vanishes $C=0$, agreeing with the separable state obtained in Eq.~\eqref{eq:phi0_state}. In the following subsections we will independently analyse the effect of each of the imperfections in more detail.

\subsection{Fine-structure splitting}
In this subsection, we analyse the effect of the FSS on the entanglement quality of the path-entangled state. Non-zero FSS leads to a spin-flip between the exciton levels ($\ket{X_\pm}$), and it is therefore convenient to describe the decay in the linear polarisation basis with $x$- and $y$-polarized states, $\ket{X_x}$ and $\ket{X_y}$ respectively (c.f. Fig.~\ref{fig:scheme}(b)). In this basis, the states are decoupled and the FSS induced spin-flip frequency $S$ corresponds to an energy splitting 
between the exciton levels. 
The splitting makes the emitted photons distinguishable in energy, and crucially their frequencies are correlated with their polarisations.
This leads to ``which-way'' information about the polarisation state, which means reduction in the degree of entanglement. To overcome this issue Ref.~\cite{fognini2018} has proposed using electro-optical modulators that rotate the polarisation of the biexciton and exciton photons separately to effectively erase the information gained from the splitting in the polarisation-encoded state. A phase modulator could similarly be applied to improve path-entangled states. Alternatively, narrow spectral filtering in between the two frequency components of either the exciton or biexciton emission can be implemented to erase the ``which-path'' information, however, at the expense of significantly reducing the entanglement generation rate \cite{akopian2006}. Another approach is to implement QDs with improved symmetry in order to obtain a smaller splitting $S$ \cite{huo2013}.

The reference situation corresponds to an ideal system without fine structure splitting and perfect directional (chiral) coupling ($S=0$ and $\Phi=\pi/2$). This situation is easily understood from the level structure in Fig.~\ref{fig:scheme}(a), where emission occurs with two oppositely polarized circular dipoles ($\sigma_-$ and $\sigma_+$). With perfect chiral coupling these decay in opposite directions creating that maximally entangled state $(\ket{AB}+\ket{BA})/\sqrt{2}$. As a consequence, the probability of detecting both photons on the same side of the waveguide vanishes 
(dashed line in Fig.~\ref{fig:S_pi_2}). The probability of detecting one photon at each of the opposite ends of the waveguide 
decays exponentially with the exciton spontaneous emission rate $(\gamma'_x+\gamma'_y)/2$ (dotted line in Fig.~\ref{fig:S_pi_2}) as expected from the lifetime of the exciton states.

We now consider a scenario where the FSS creates an asymmetry between the exciton levels ($S\neq 0$), while the chiral coupling is still ideal ($\Phi = \pi/2$). This generates an oscillation between two maximally entangled states as discussed below Eq.~\eqref{eq:entanglement_oscillation}. The corresponding probabilities of the various detection patterns is shown with the dash-dotted and solid lines in Fig.~\ref{fig:S_pi_2}. The amplitude of oscillations decays exponentially with the time constant set by the exciton spontaneous emission rate. As discussed in the previous section, although the emitted state changes over time, 
it remains maximally entangled, i.e., $C(\tau\geq0)=1$, and it is a superposition of standard Bell states.

\subsection{Imperfect chirality}
We now analyse the joint effect of both imperfect chirality ($\Phi\neq\pi/2$) and non-zero FSS ($S\neq 0$). An example of the detection probability for this situation is shown in Fig.~\ref{fig:3ab}(a). Curiously, the probabilities $P_{AA}$ and $P_{BB}$ are out of phase, meaning that with a given time delay there is a difference in the probabilities of detecting two photons at the two ends of the waveguide. This effect happens due to an interplay of the imperfect chirality and the FSS. A decay from the biexciton state and subsequent detection of the photon at one end creates a coherent superposition between the two exciton states $\ket{X_x}$ and $\ket{X_y}$ with a phase $\mp\Phi$ depending on where the photon was detected. The subsequent dynamics induced by the FSS $S$ may then evolve the state towards or away from the relative phase $\pm \Phi$, which gives the maximal emission in the same direction.

With non-perfect chirality, the concurrence $C$ of the path-entangled, bi-photon state emitted by the biexciton cascade is reduced since the imperfect chirality limits the directional coupling of the QD emission.
The dependence of $C(\tau)$ on $\Phi$ is shown in Fig.~\ref{fig:3ab}(b).
We observe that $C$ is independent of $\tau$ only if $\Phi = n\pi/2$, where $n$ is a non-zero integer. 
If $n$ is even, $C(\tau\geq0) = 0$ and corresponds to the completely non-chiral case. If $n$ is odd, we reproduce the results of the perfect chiral case that results in a maximally entangled state with $C(\tau\geq0) = 1$ as discussed in the previous subsection. 
For partial chirality $\Phi \neq \pi/2$, the FSS induces oscillations between non-maximally entangled states and $C$ oscillates as a function of the detection time $\tau$. 
In general, $C$ is below unity except for $S\tau=\pi$, where the concurrence is unity for all $\Phi\neq0,\pi$. 

\subsection{Timing jitter}\label{sec:TimeJitter}
\begin{figure}
\centering
\includegraphics[width = \linewidth]{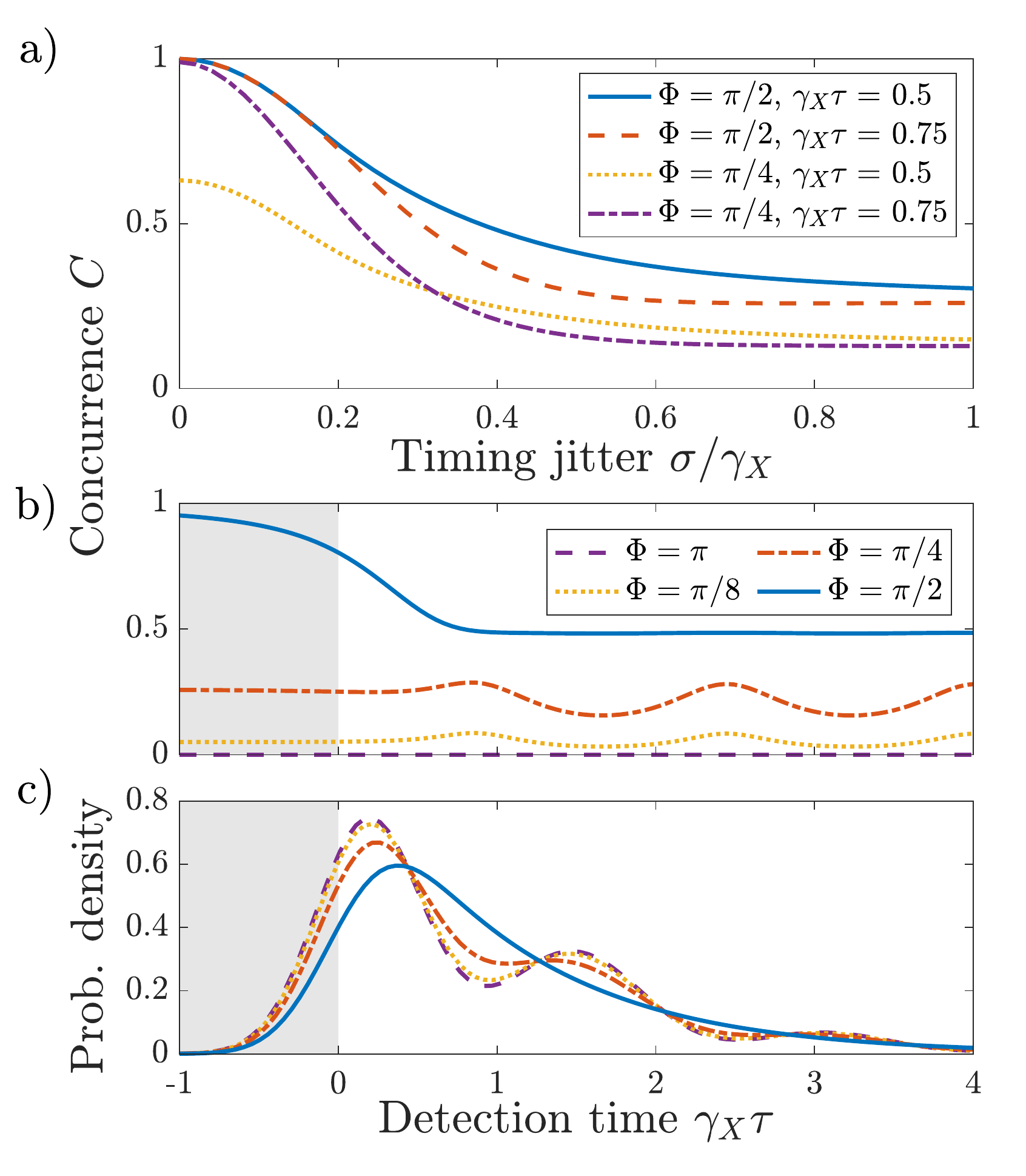}
\caption{\label{fig:time_jitter} a) Concurrence $C$ of the state as a function of the detection timing jitter, quantified by the Gaussian RMS width $\sigma$, for several values of chirality $\Phi$ and time difference $\tau$ (the selected values are marked with $\times$-symbols in Fig.~\ref{fig:3ab}(b)). b) Concurrence $C$ of the state for different degree of chirality, quantified by the phase $\Phi$, with a timing jitter of $\sigma=0.3/\gamma_X$. 
The concurrence can be larger 
at negative than positive time intervals (shaded region) since such events effectively have less timing uncertainty than those with positive time intervals (see main text).
c) Corresponding probability density $\bar{N}$ of the detection time for the situation analysed in (b). Note that the probability density quickly approaches zero for $\tau<0$ (shaded region) in contrast to the increase in concurrence. Throughout the figure we assume a non-zero FSS of $S=4\gamma_X$.}
\end{figure}

In this subsection we analyse the effect of uncertainty in the timing of photodetection events on the entanglement quality. We model the uncertainty in detection time by averaging the density matrix \eqref{eq:density_matrix} elements $\rho_{n,n',m,m'}$ with a Gaussian probability distribution with standard deviation $\sigma$ 
\begin{equation}
\begin{split}
  \bar{\rho}_{n,n',m,m'}(\tau) = \int_{0}^{\infty}d\tau' \exp\left[{-\frac{(\tau'-\tau)^2}{2\sigma^2}}\right] \\ \times \psi_{n,m}(\tau')\psi^{*}_{n',m'}(\tau')\,.
  \label{eq:concurrence_1}
  \end{split}
\end{equation}
The time-averaged density matrix $\bar{\rho}(\tau)$ is then given by
\begin{equation}
  \bar{\rho}(\tau) = \frac{1}{\bar{N}(\tau)}\begin{pmatrix} 
  \bar{\rho}_{AAAA}(\tau) & \bar{\rho}_{AAAB}(\tau) & \dots & \bar{\rho}_{AABB}(\tau) \\
  \bar{\rho}_{ABAA}(\tau) & \ddots & & \vdots\\
  \vdots & &\ddots& \vdots \\
  \bar{\rho}_{BBAA}(\tau) &  \dots   & \dots & \bar{\rho}_{BBBB}(\tau) 
  \end{pmatrix}\,,\label{eq:rho_avg}
\end{equation}
where $\bar{N} = \int_{-\infty}^{\infty}d\tau' \exp[-(\tau'-\tau)^2/(2\sigma^2)]N$ is a normalisation constant equal to the probability density of the detection time and $N$ 
is given by Eq.~\eqref{eq:norm}. From this density matrix we can then calculate the concurrence $C$.

Figure~\ref{fig:time_jitter}(a) shows the dependence of the concurrence $C$ 
on the detection timing jitter $\sigma$ at different combinations of chirality and time delay. As seen in the figure the concurrence drops when the uncertainty in detection time becomes comparable to the oscillation period $1/S$.
This highlights 
the importance of keeping track of the time dependence for the quality of the final path-entangled state.
Unlike the jitter-free case, even systems with perfect chirality ($\Phi = \pi/2$) exhibit $C<1$ for non-zero values of $\sigma$ since we do not know precisely which state we have. 
The asymptote of $C$ with increasing time jitter is observed to depend only on the phase $\Phi$, i.e., when the time jitter is comparable to or larger than the spread in emission time; the precise time of the detection is not important.
Figure~\ref{fig:time_jitter}(b) shows the time evolution of $C$ for a fixed timing jitter $\sigma = 0.3/\gamma_X$ for different values of the chiral phase and $S=4\gamma_X$.
Note that a peculiar effect occurs for $\Phi=\pi/2$ (Fig.~\ref{fig:time_jitter}(b)), where we observe that $C$ increases at negative time delays (grey shaded region). 
Since the emission of the exciton always occurs after the biexciton emission ($t_{X}-t_{XX}=\tau>0$), negative detection intervals ($\tau<0$) correspond to the case where the emission of the photon must have occurred close to $\tau=0$, i.e. with minimal time delay, but was measured to be at a negative value due to the time jitter. 
Therefore the uncertainty in the emission time, which is otherwise given by the detection time jitter, is effectively reduced for negative detection times, leading to a higher concurrence.
The probability of measuring the state at negative time intervals, however, decays very rapidly as $\tau$ decreases, as shown in Fig.~\ref{fig:time_jitter}(c).
The larger concurrence at small positive time delays ($0<\tau\lesssim\sigma$) compared to later times 
can be understood with similar arguments.
On the other hand for $\Phi=\pi/4 $ and $\Phi=\pi/8$ the fidelity in the absence of time jitter is lower around $\tau=0$ than at later times, c.f. Fig.~\ref{fig:3ab}(b). As a consequence the peak concurrence still occurs around $S\tau=\pi$.
The probability density in Fig.~\ref{fig:time_jitter}(c) decays with the decay rate $\gamma_X$ of the exciton states. On top of this it oscillates with increasing amplitude as the system becomes less chiral ($\Phi\rightarrow 0$). The reason is that the polarisation of waveguide modes becomes linear as the system loses chirality. After the decay of the biexciton, the polarisation of the exciton state rotates due to the FSS $S$ and may thus be more or less aligned with the waveguide polarisation. In contrast, for the chiral case the waveguide polarisation is circular and the rotation of the polarisation does not affect the decay rate.

\subsection{Asymmetric exciton decay}
\label{sec:asym}
\begin{figure*}
\centering
\includegraphics[width = \linewidth]{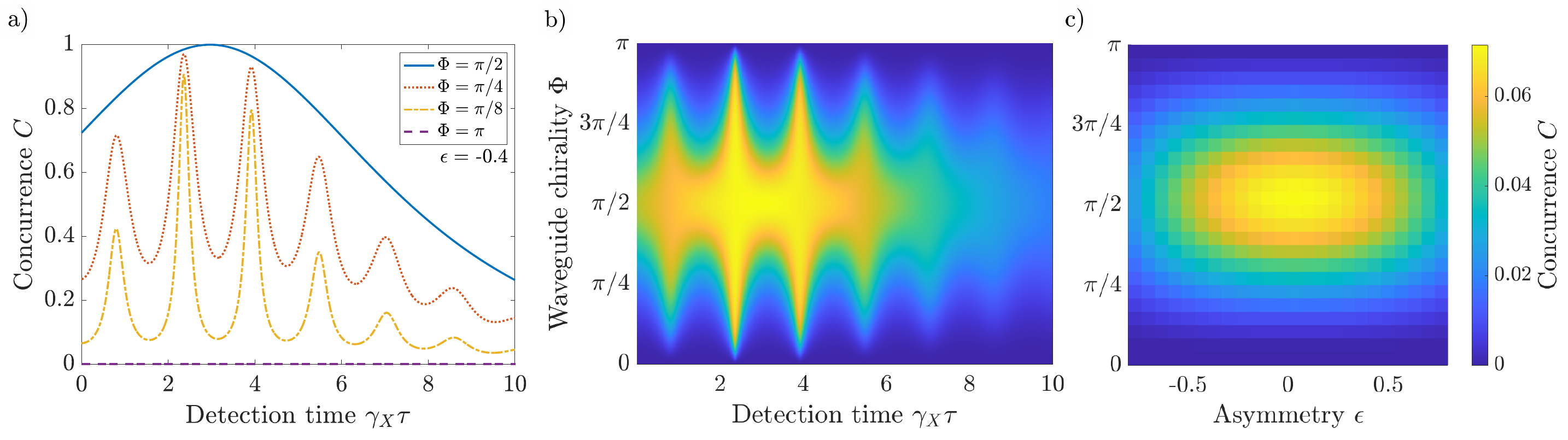}
\caption{\label{fig:4abc} a) Concurrence $C$ of the state as a function of the time delay $\tau$ at different degrees of the chirality, quantified by $\Phi$. The asymmetry between the exciton decay rates is described by the parameter $\epsilon=(\gamma_x-\gamma_y)/(\gamma_x+\gamma_y)$ and fixed to be $\epsilon=-0.4$. The oscillations in the concurrence due to the FSS $S=4\gamma_X$ are modulated by the difference in exciton decay rates. b) Colour map of the concurrence of the state as a function of the difference in phase $\Phi$ and the time $\tau$ for $\epsilon=-0.4$. The concurrence oscillates in time due to the FSS $S$, and reaches a maximum for a non-zero time $\tau$. This optimal difference in emission time is reached when the difference in populations of the exciton states arising from a difference in biexciton decay rates is cancelled by the faster decay of the most likely state. c) Colour map of the concurrence $\bar{C}$ averaged over all emission times $\tau$ as a function of the difference in phase $\Phi$ and the asymmetry parameter $\epsilon$ with $S=4\gamma_x$ and a time jitter of $\sigma=3/\gamma_X$. We observe that the concurrence is optimal for perfectly symmetric exciton decays, as expected.}
\end{figure*}

In the experiments presented in Ref.~\cite{freja}, the decay rates of the $x$ and $y$-polarized exciton levels were nearly identical (i.e. $\gamma_x\approx\gamma_y$). However, in general these two decay rates may differ depending on the position of the QD in the waveguide, with the asymmetry more dominant at locations with a low degree of directional emission, i.e., far from perfect chirality. In this subsection we analyse how this asymmetry can affect the quality of entanglement.

Figure~\ref{fig:4abc}(a,b) shows the impact of asymmetry $\epsilon\equiv(\gamma_x-\gamma_y)(\gamma_x+\gamma_y)$ on the concurrence for the case of $\epsilon = -0.4$. 
As the decay rates of the $x$- and the $y$-polarized exciton levels are different, one can gain ''which-path'' information about the photon decay from the photodetection time, i.e. the highest decay rate would result in increased likelihood of early detection of photon, and vice versa. 
This extra information about the emission process reduces the entanglement. Furthermore, the difference in decay rates of the biexciton state creates a difference in populations of the $\ket{X_x}$ and $\ket{X_y} $ states. However, if the difference in the emission time is comparable to the 
difference in decay rates, the `which-path' information arising from the asymmetric decay rates is erased and 
the entanglement is recovered. 
This interplay between the difference in emission times and the asymmetry $\epsilon$ leads to an optimal time delay $\tau$ that maximizes the concurrence as observed in Fig.~\ref{fig:4abc}(a,b).
In addition to this optimality, we still observe that $C$ oscillates with emission time delay 
due to the non-zero $S$ as discussed in Sec. III.A.

For a systematic study of the effect of asymmetry, 
we calculate the average concurrence $\Bar{C}$ over all $\tau$ detection times, defined as
\begin{equation}
  \Bar{C} = \int_{-\infty}^{\infty}P(\tau)C(\tau)d\tau\,,
\end{equation}
where $P(\tau)$ is the corresponding probability density of the state at time $\tau$. 
The dependence of $\Bar{C}$ on the asymmetry parameter $\epsilon$ and the phase difference $\Phi$ is shown in Fig.~\ref{fig:4abc}(c), which highlights that $\Bar{C}$ is maximized for symmetric decay of the exciton dipoles, i.e. $\epsilon = 0$.

\subsection{Dephasing noise}

Electron-phonon interactions can induce dephasing processes that will degrade the indistinguishability of photons. These processes are nevertheless not expected to affect the entanglement quality of the state. This is due to the two exciton levels being symmetrically perturbed by the phononic interaction: the dephasing of the exciton level is expected to be induced solely by deformation potential of the quantum dot, which is independent of its spin properties \cite{muljarov2004a,tighineanu2018}, so that the two levels are dephased in an identical manner. Therefore the indistinguishability of the photons emitted at the exciton level is reduced by this effect, but it is not expected to degrade the entanglement quality, as witnessed experimentally in Ref.~\cite{coste2023}.

\section{Conclusion}
We have provided an in-depth analysis of the entanglement properties of a QD biexciton cascade embedded in a chiral nanophotonic waveguide, as experimentally realised in Ref.~\cite{freja}. We have calculated how the biexciton cascade can deterministically prepare a path-encoded state mediated by the chiral-coupling of the waveguide. The entanglement of the state is, however, affected by errors unavoidably present in the experimental implementation of the system. In particular, we have shown how the time dependence of the state induced by the FSS plays a crucial role in determining the generated entanglement. 
The amount of path-entanglement generated by the biexciton cascade can strongly depend on the emission time, 
while the presence of detection time jitter reduces the concurrence of the state.
Finally, imperfect directional-coupling in the waveguide reduce the concurrence of the path-encoded entangled state as well. Our work quantifies the role of such imperfections and lay out a route to a deterministic source of path-encoded entangled photons of high entanglement quality. We hope our work will motivate further experimental improvements of this novel entanglement source. 

\section{Acknowledgments}
We acknowledge the support of Danmarks Grundforskningsfond (DNRF 139, Hy-Q Center for Hybrid Quantum Networks).

\bibliographystyle{apsrev4-1}

\bibliography{mybib_biexciton}
\end{document}